# COMMON METRICS TO BENCHMARK HUMAN-MACHINE TEAMS (HMT): A REVIEW

**PRAVEEN DAMACHARLA[1] (Graduate Student Member, IEEE), AHMAD Y. JAVAID[1] (Member, IEEE), JENNIE J. GALLIMORE[2] (Senior Member, IEEE), AND VIJAY K. DEVABHAKTUNI[1] (Senior Member, IEEE)**

[1]Department of Electrical Engineering and Computer Science, the University of Toledo, OH 43606, USA
[2]Department of Biomedical, Industrial, and Human Factors Engineering, Wright State University, Dayton, OH 45435, USA

Corresponding author: Ahmad J. Javaid (email: Ahmad.Javaid@utoledo.edu).

This work was supported in part by the University of Toledo and Round 1 Award from the Ohio Federal Research Jobs Commission through Ohio Federal Research Network (OFRN)

**ABSTRACT** A significant amount of work is invested in human-machine teaming (HMT) across multiple fields. Accurately and effectively measuring system performance of an HMT is crucial for moving the design of these systems forward. Metrics are the enabling tools to devise a benchmark in any system and serve as an evaluation platform for assessing the performance, along with the verification and validation, of a system. Currently, there is no agreed-upon set of benchmark metrics for developing HMT systems. Therefore, identification and classification of common metrics are imperative to create a benchmark in the HMT field. The key focus of this review is to conduct a detailed survey aimed at identification of metrics employed in different segments of HMT and to determine the common metrics that can be used in the future to benchmark HMTs. We have organized this review as follows: identification of metrics used in now, and classification based on functionality and measuring techniques. Additionally, we have also attempted to analyze all the identified metrics in detail while classifying them as theoretical, applied, real-time, non–real-time, measurable, and observable metrics. We conclude this review with a detailed analysis of the identified common metrics along with their usage to benchmark HMTs.

## I. INTRODUCTION

The future of technology lies in human-machine collaboration rather than on a completely autonomous artificial intelligence (AI). Dr. Jim Overholt, senior scientist at the Air Force Research Lab (AFRL), stated, "The US Air Force Research Laboratory (AFRL) has no intention of completely replacing humans with unmanned autonomous systems" [1]. Therefore, to achieve the best results, a human-machine teaming or collaboration is the only choice we have, but such a teaming comes with its own set of challenges. We propose to define HMT as a combination of cognitive, computer, and data sciences; embedded systems; phenomenology; psychology; robotics; sociology and social psychology; speech-language pathology; and visualization, aimed at maximizing team performance in critical missions where a human and machine are sharing a common set of goals. Team members will share tasks, and the machine may provide suggestions that can play a crucial role in team decision-making. Such a collaboration requires a two-way flow of information. Based on the above-proposed definition, to be deemed as an HMT, a team should contain at least one human and one machine/intelligent system. Perhaps the best example of practical use of an HMT can be attributed to a 2005 game of chess. In this game, two inexperienced chess players teamed up together with three PCs and won a chess competition against a group of supercomputers and grandmasters, which did not form a team. In this scenario, human team members were able to leverage the machine's data mining and information processing capabilities based on their cognition skills [2]. Although machines have been used to assist humans for decades, these systems are not collaborative partners but are programmed for specific tasks [3]. The primary concern of HMT is effective integration of human and machine tasks so that the team collaboration optimizes the efficiency of critical tasks [4-8]. Making successful outcomes consistent and repeatable with high accuracy would also demonstrate an effective HMT that is possible only through comprehensive studies.

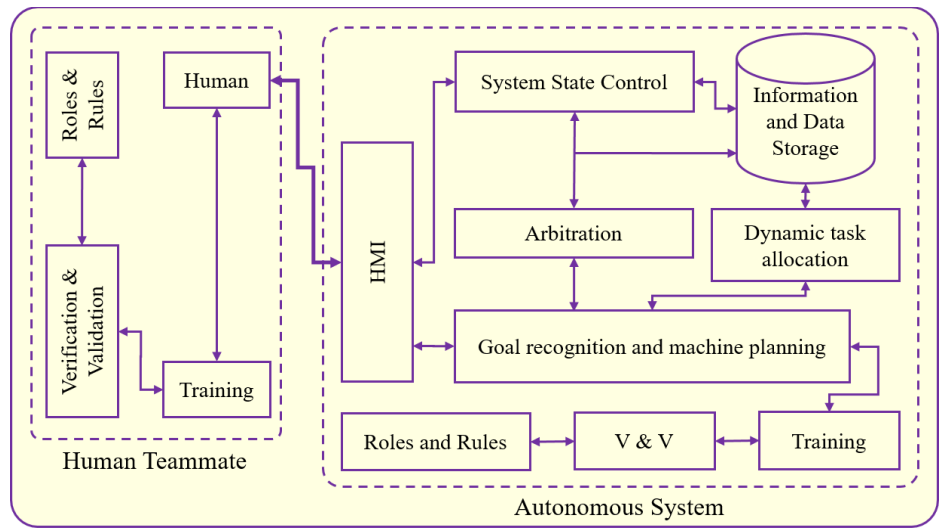

**Figure 1.** Generalized Human Machine Team Model

### A. HMT Overview

By analyzing various published works [8-21], we identified six major HMT components, with architectures, interfaces, and metrics being the highly researched areas. We present brief definitions and examples of these components:

1) ARCHITECTURES: The founding principle of building an HMT architecture is to achieve an optimal machine assistance. Architecture is necessary to set boundaries, assign duties, and design interfaces to increase the team effectiveness. Through analysis of 19 published frameworks, we identified nine essential functional blocks for a generic HMT framework: human-machine interaction (HMI), information and data storage, system state control, arbitration, goal recognition and mission planning, dynamic task allocation, rules and roles, verification and validation (V&V), and training [21-35]. This is shown in figure 1.

2) INTERFACES: Any focus over interface and interaction method will enable an effective human-machine communication. The association for computing machinery defines HMI as "a discipline concerned with the design, evaluation, and implementation of interactive computing systems for human use and with the study of major phenomena surrounding them" [36]. The HMI can be divided into three principal components: the user, the interface, and the machine. Here, an interface is a device that typically encompasses both software and hardware to streamline an interaction between user and machine. Examples include a graphical user interface, web browsers, and various I/O devices [37]. Many published studies that have classified and analyzed interfaces used in HMI are acknowledged in [38-40].

3) METRICS: Metrics are crucial measures to track, assess, and compare a process, task, or system with respect to performance, usability, efficiency, quality, and reliability as defined by the system performance goals. Metrics can also be used to evaluate the effectiveness of an HMT and its agents (human, machine, and team) on various levels.

4) ROLES AND RULES: Roles are defined as assumed or assigned responsibilities within a system, process, or task. On the other hand, rules are defined as a set of explicit regulations governing conduct in a situation or activity. By analyzing published work, we concluded that requisite and opportunistic are two categories of roles and rules. Implementing roles and rules in HMT helps generate a symbiotic human-machine ecosystem that will think as no human has ever thought and will process the data in a way that no machine ever processed [4, 31, 41-45].

5) TEAM BUILDING: According to the earlier works of researchers presented in [46], teams are defined not as just individual parts of machinery but they must be built together. In an HMT, one can build a systemic team with compatible team members. Through literature review, we identified that team development has two dimensions: (1) the task dimension consisting of forming, conflict resolution, norming, and performing, and (2) the interpersonal dimension consisting of dependency, conflict, cohesion, and interdependence [46-53].

6) VERIFICATION AND VALIDATION (V&V): For a team to function optimally, features such as trust, cohesion, expectations, and motivation must be considered because of their effects on team performance. V&V is a crucial component of HMT that helps validate the team-building features mentioned above and thus gives key insights for optimizing the team formation and performance. The V&V methods can be further classified into two groups based on their use: during mission and training [18, 54-59].

### B. METRICS BACKGROUND

Although the foundations of HMT were laid at Defense Advanced Research Projects Agency in 2001 [10], it took another five more years for the research community to identify a set of metrics that facilitates a well-organized structure of

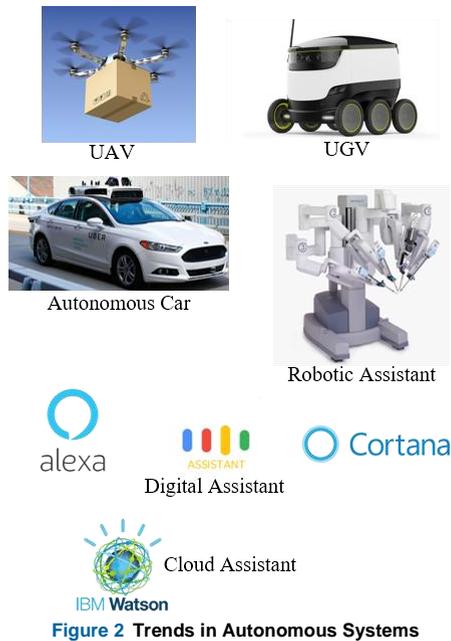

UAV

UGV

Autonomous Car

Robotic Assistant

alexa

Digital Assistant

Cortana

Cloud Assistant

IBM Watson

**Figure 2** Trends in Autonomous Systems

TABLE I: KEYWORDS USED

| Primary Keywords | |
|---|---|
| Core Concepts | Keywords |
| Human Machine Teaming | Control metrics, interface metrics, synthetic assistant, synthetic mentor, intelligent assistant, rules and roles, symbiosis, verification and validation, measuring methods, physiological attributes |
| Human Machine Collaboration | Metrics, architecture, interface, team building, human factors, ergonomics, task, automation, shared control, symbiosis, physiological attributes |
| Human in Team | Team building, metrics, interface, human factors, human-robot collaboration, ergonomics, multirobot controls, shared control, physiological attributes |
| Machine in Team | Robot control, software agent, synthetic assistant, metrics, synthetic mentor, intelligent assistant, team building, interface, human factors, multirobot teams |
| Metrics | List of all identified metrics in Table IX + measuring methods |
| **Secondary Keywords** | |
| UAV, UGV, navigation, surveillance, healthcare, medical assistant, identification, ordnance disposal, geology | |

human-robot interactions (HRI). For various metrics, we found close but different descriptions of the same metric, primarily for various HMIs in human-robot or robot-only swarms [20, 60]. The research community uses metrics that are application and domain specific. For example, researchers in [61] developed an approach to define human supervisory control metrics while [62] has identified common metrics for HRI standardization. Researchers in [63, 64] focused on developing false alarm metrics to analyze erroneous HRI. The robot performance evaluation metrics for understanding team effectiveness are defined in [65]. Researchers have also developed metrics from human-computer interaction (HCI) heuristics to aid information analysis in interactive visualizations [66]. This work made an active effort to define metrics for specific components of HMT, such as HCI, HRI, and architectures, whereas research on common metrics is limited.

Identifying common metrics will allow benchmarking of HMT designs, comparison of findings, and development of evaluation tools. The primary difficulty in defining common metrics is the diverse range of HMT applications. In this review, we focus on metrics for all three agents of HMT, for example, human, machine, and the team. The goals of this review paper are (1) identification and classification of metrics, (2) evaluation of the identified metrics to find common metrics, and (3) proposal of common metrics that can be used in future HMT benchmarking. The rest of the paper is structured as shown in figure 3.

## II. METHOD

### A. KEYWORDS AND DATABASES

To limit the scope of this study, we developed a set of keywords based on pertinent technological and scientific domains that focus on HMT. The HMTs investigated in this study account for one or more task-oriented mobile robots or software agents as machine team member(s) and at least one human as a team member. Further, the machines that take part in an HMT must belong to one of the following categories: unmanned aerial vehicles (UAVs), unmanned ground vehicles, AI robots, digital assistants, and cloud assistants, as shown in figure 2. The search was limited to HMT applications in target search and identification, navigation, ordinance disposal, geology, surveillance, and healthcare. The keywords used are listed in Table I and the databases utilized are as follows: IEEE Xplore, Science Direct (SCOPUS/Elsevier), Defense Technical Information Center, SAGE Publications, and Google Scholar.

### B. SELECTION CRITERIA

The following criteria were set to evaluate the articles found after a detailed search. Firstly, we tried to define the relevance of the article with our objectives/goals as follows:
- Discusses HMT or human-machine collaboration?
- Discusses one or more HMT components?
- Discusses metrics related to an HMT agent?
- Mentions or discusses core HMT concepts?

Articles that satisfied the above criteria were further filtered based on primary and secondary keywords used in specific sections. Further, we identified metrics that relate to teaming and HMT and conducted another refined search to obtain the most relevant literature. Out of hundreds of articles identified in the search process, a total of 188 articles were considered for the review.

### C. LIMITATIONS

A key limitation of this review is the breadth of the review since the area of HMT is extensive and involves many fields of study. For the sake of this review, a limited number of primary articles are reviewed here (n=77). Such a wide-range review poses a bigger challenge in terms of comprehensive coverage of various metrics and related research questions. Therefore, the review focuses on three agents of HMT that are worthy for an in-depth review. We selected the most relevant information available from the literature. Another limitation that entails establishing common metrics for all HMT types or benchmarking them on a single scale is the dependence on many factors such as application, and the number of agents.

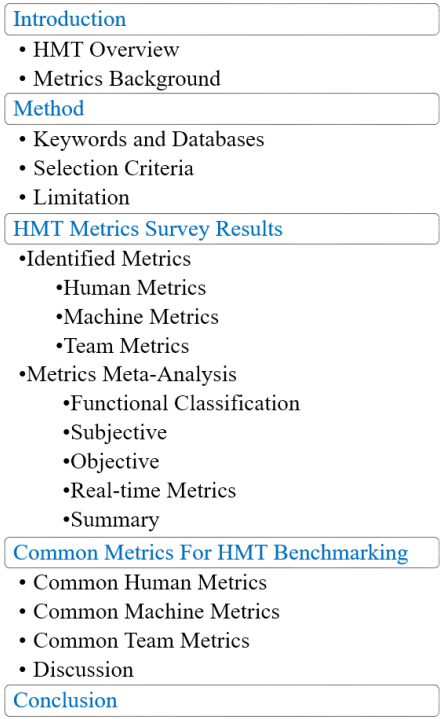

**Figure 3** Review paper structure

## III. HMT METRICS SURVEY RESULTS
In this section, we present a comprehensive and classified metric list for the three agents of HMT: human, machine, and team (or system). This strategy resulted in (1) an analysis that applies to a specific range of applications, and (2) the ability to assess the application specific HMT performance.

### A. IDENTIFIED METRICS
#### 1) HUMAN METRICS
This subsection identifies metrics that measure different human aspects such as system knowledge, performance, and efficiency that can be used to evaluate a human agent in an HMT. Most of the metrics we present in this section are well established by various scientific studies.

*Situational awareness (SA)* is measured by monitoring task progress and sensitivity to task dynamics during execution. The degree of *mental computation* estimates the amount of *cognitive workload* an operator manages to complete a task, for example, a task that requires object reference association in *working memory* a or user's cognitive abilities to perceive projections of the real-time environment [62, 67]. The accuracy of a *mental model* of an operator depends on interface comprehensiveness and simplicity in addition to control and compatibility a machine provides. *Attention allocation* measures the attention an operator pays to a team's mission and the operator's ability to assign strategies and priorities of tasks dynamically. The metric also considers an operator's degree of attention over multiple agents. It is measured using eye tracking, duration of eye fixations to an area of interest, and task completion rate, while *attention allocation efficiency* is measured using wait times [61, 68, 69]. *Intervention frequency* is the frequency with which an operator interacts with the machine [20]. As per literature, operators' intervention frequency is also known as *intervention rate* or *percentage requests*. *Stress* can be physical or mental. However, both may indicate the operator workload and are measured in two ways. First, researchers perform sample testing of humans' stress hormones, such as hypothalamic-pituitary adrenaline, cortisol, and catecholamine, which are found in blood, saliva, and urine samples [70]. Second, researchers can perform a detrended fluctuation analysis of a human's heartbeat [71].

Human *safety metrics* involve evaluation of the risk posed to the human life while working near machines, for example, the location of the machine relative to the human. These mostly apply to applications in a high-risk environment such as threat neutralization. Human factor studies suggest that humans can establish the best cooperation with a machine through a 3D immersive environment [72]. In [73, 74], researchers suggest that humans can be more effective when the environment and goals are in their best interest.

Other human performance attributes such as *psychomotor processing*, *spatial processing*, *composure*, and *perseverance* are important to improve the team cohesion through human performance enhancement. Overall personal (physiological, cognitive, and psychological) attributes have been classified into five subdomains after a detailed study by several defense agencies and are summarized in Table II [75-78].

TABLE II: HUMAN PERFORMANCE METRICS FOR HMT

| Sub Domain | Attributes |
|---|---|
| *Physical Health* | general health; stamina; stress; fatigue |
| *Cognitive* | cognitive proficiency; attention; spatial processing; |
| *Perception* | memory; psychomotor processing; reasoning |
| *Intra-Personal* | composure; resilience; self-certainty; conscientiousness; success oriented; perseverance; decisiveness; impulsiveness; cohesiveness; assertiveness; adaptability; self-confidence |
| *Inter-Personal* | extraversion; judgment; team oriented; adaptability |
| *Motive* | moral interest; occupational interest |

#### 2) MACHINE METRICS
All the machine-level metrics related to HMT, such as efficiency, performance, and accuracy, are well represented in literature. A few more are detailed as follows: *machine self-awareness*, or the degree to which a machine is aware of itself (limitations, capacities), is a precursor to reducing the human cognitive load and measured based on *autonomous operation time*, the *degree of autonomy*, and *task success* [62, 79]. Technically, *unscheduled manual* operation time may either be an interruption period in current plan execution or an unexpected assigned task [80]. *Neglect tolerance* (NT) is interpreted in numerous ways, such as machine performance falling below expectation, time to catch-up, the idle period, or operation time without user intervention. *State metric* helps track the machine or plan state based on four dynamic states: assigned, executed, idle, and out of the plan. *Robot attention demand* (RAD) is a measure of the fractional "task time" a human spends to interact with a machine. *Fan out (FO)* is a measure of how many robots with similar capabilities a user can interact with simultaneously and efficiently and is inverse of RAD [81]. *Interaction effort* (IE) is a measure of the time required to interact with the robot based on experimental values of NT and FO and is used to calculate RAD [81, 82]. Although humans can communicate through visual cues, gestures, etc., most machines need accurate information to act. Such information is mostly sent over wireless channels for various cyber-physical or cloud-robotics systems such as UAVs [83, 84]. Studies suggest that communication with machines in real time can be accomplished successfully by adapting to 5G communication technologies in hardware and software implementation [85-87].

Additional machine metrics that do not have a quantitative representation yet, and are difficult to measure, include *resource depletion, subgroup size, collision count, usability, adequacy, sensory-motor coefficient, level of autonomy discrepancies, physical constraints,* and *intellectual constraints* [20, 61, 80].

### 3) TEAM METRICS

Conventionally, a *team* has two primary components: a leader and one or more team members. A team leader is someone who provides guidance and instruction and leads the group to achieve set goals. In contrast, a team member is an individual who works under the supervision of a team leader [88]. Although there is no quantitative validation or representation of a team member, many guidelines and studies define the characteristics of an ideal team member who serves as a reference to evaluate or prepare a machine as a team member. Five essential features of a human team member are defined as follows: functional expertise, teamwork, communication skills, job assignment flexibility, and personality traits [89]. In contrast, a well-defined or established machine team member feature list seems to have not been researched well due to the nascent nature of HMT research.

The key focus of team metrics is mission assignment and execution. *Task difficulty* represents the mental load a particular task generates [90]. The task difficulty metric for a machine depends on FO and requires three factors for measurement: *recognition accuracy*, *situation coverage*, and *critical time ratio* of a machine [65]. *Recognition accuracy* is the ability of the machine to sense its I/O parameters. *Situation coverage* (SC) is the percentage of situations encountered and accomplished by the robot. SC is defined based on *plan* and *act* stages of the mission. *Critical time ratio* is the ratio of time spent by a robot in a critical situation to the total time of interaction [65]. *Network efficiency* is the rate of flow of information between the human and the machine and determines the efficiency of interaction. It also influences time taken for scheduled and unscheduled manual operations, accuracy of mental computation, negligence tolerance, and human-machine ratio [20]. Four well-known subclasses of *false alarms* are true positive (TP), true negative (TN), false positive (FP), and false negative (FN) [63]. While false alarms measure complex communication between humans and machines in a team, people may ignore false alarms. A human factor study presented a trade-off between ignoring false alarms and *misses* and concluded that alarms are strongly situation dependent [91]. Some other team metrics that can be used in effective interactions are *hits*, *misses*, *automation bias, and misuse of automation* or metrics based on application scenario [92]. *Robustness* measures the ability of the team to adapt to the changes in task and environment during task execution [93] while *productivity* measures productive time compared to total invested time. *Task success ratio* indicates the number of completed versus allocated tasks [80].

Additional team metrics include *team effectiveness, human-robot ratio, cohesion, neighbor overlap, total coverage, critical hazard, autonomy discrepancies, TP, TN, FP, and FN interaction rates (TPIR, TNIR, FPIR, FNIR), cognitive interaction, cryptic coefficient* and *degree of monotonicity* [20, 94].

### B. METRICS META-ANALYSIS

To identify common metrics, we need to analyze the metrics for properties such as aspect of measure, measurement technique, reliability and dependability of measurements, performance, and suitability for selected application area.

TABLE III: FUNCTIONAL CLASSIFICATION OF HMT METRICS

| Functionality | List of Metrics |
|---|---|
| Efficiency metrics | attention allocation; decision accuracy; mental workload; mental computation; workload; mental models; usability; sensory motor coefficient; plan execution; interaction efficiency; monotonicity; effort; cryptic coefficient; network efficiency; accuracy and coherence of mental models; recognition accuracy; fan out; span of control; flexibility; level of autonomy discrepancies; false alarms; true positive interaction rate false positive interaction rate; true negative interaction rate; false negative interaction rate; collision count; percentage request by operator; percentage request by machine; mode error; team productivity |
| Timing metrics | neglect tolerance; critical time ratio; autonomous operations time; manual operation time; scheduled operation time; unscheduled operation time; completion time; execution time; productive time; team performance; task success; intervention response time; intervention frequency; mutual delay; settling time; operator to robot time ratio; Mean Time Between Interventions (MTBI); Mean Time Completing an Intervention (MTCI); Mean Time Between Failures (MTBF) |
| Mission metrics | reliability; trust; total coverage; task allocation; plan state; plan execution; plan idle; plan out; neighbor overlap; similarity; task difficulty; situation coverage; robot attention demand; resource depletion and task success |
| Safety metrics | Risk to human; general health; critical hazard; fatigue; stress; self-awareness; human awareness; situation awareness |

TABLE IV: APPLIED METRICS

| Human | Machine | Team |
|---|---|---|
| Adaptability | Usability | Cohesion |
| Assertiveness | Fan Out | Interventions |
| Impulsiveness | Robot Attention Demand | Intervention Response Time |
| Cohesiveness | Collision Count | Neglect Tolerance |
| Perseverance | Plan Execution | Unscheduled Operations Time |
| Extraversion | Plan Idle | Time Autonomous Operations |
| Conscientiousness | Plan out | Time in Manual Operations |
| Humility | Plan State | Plan State |
| Occupational Interest | Resource Depletion | Situation Coverage |
| Psychomotor processing | Interaction Effort | Task success |
| Stamina | Mutual Delay Time | Task Difficulty |
| General health | Neglect Tolerance | False Alarms |
| Fatigue | Settling Time | False Positive Interaction Rate |
| Stress | Time in Autonomous Operations | False Negative Interaction Rate |
| Situation Awareness | Time in Manual Operations | Interaction Efficiency |
| Attention Allocation Efficiency | Unscheduled Operations Time | Network Efficiency |
| | | Recognition Accuracy |
| | | Team Productivity |
| | | True Negative Interaction Rate |
| | | True Positive Interaction Rate |

☐ Research Metrics    ☐ Non-Research Metrics

These characteristics are identified through meta-analysis[1]. Metrics can be primarily classified based on either the measurement technique (subjective, objective, direct, indirect, nominal, ordinal, interval, ratio, process, resources, and results), or the quantity they measure (efficiency, safety, cognition, and time) [95, 96]. Here, we analyze the identified metrics based on measurement techniques, reliability, and performance and classify them as functional, subjective, objective, and real-time.

### 1) FUNCTIONAL CLASSIFICATION
Through this review, we found that several identified metrics can be employed in all three HMT agents with subtle modifications in measurement techniques; for example, the time taken by a human to complete a task can be measured using an external observer[2]. In contrast, machines use an automatic timer for the same purpose. We identified efficiency, time, mission, and safety as four functional classes of HMT metrics, as shown in Table III. Metrics to evaluate efficiency will give the observer the required V&V to tune each agent to operate with maximum efficiency [20, 62, 63, 80, 81, 93, 97, 98]. Time metrics provide data related to the time taken for different operations by machine, human, and team, and these metrics are very important in *decision-making* and *performance* and *status determination* [20, 62, 65, 80, 81, 99-101]. Mission metrics measure attributes related to a task such as planning [20, 65, 80, 81]. Safety of the team is the highest priority for any mission, especially in stochastic and dynamic mission environments. Safety metrics measure the agent and mission safety during task execution [64, 71, 72]. Another class of metrics, termed as applied metrics, deals with the practicality and research on metrics and is divided into research and non-research metrics. Table IV classifies the applied metrics with respect to the HMT agents.

### 2) SUBJECTIVE
Subjective metrics (SM) are used to measure abstract qualities based on human perception. These metrics may include feedback or judgment from observers (superiors or experienced professionals), for example, self-feedback, evaluation, or ratings.

Table V summarizes a few available well-documented SM scales. *Adaptability* is measured using a five-scale rating from the experts [102]. *Assertiveness* is measured based on the Rathus assertiveness scale [103, 104], while *resilience, composure,* and *self-confidence* are measured using 19 different scales, such as the Connor-Davidson resilience scale, student motivation scale, and resilience scale for adults [105, 106]. *Conscientiousness* is computed using the Chernyshenko scale, which is a 60-item question inventory, with each question rated by subjects on a 4-point scale [107].

TABLE V: SCALES FOR SUBJECTIVE METRICS

| Scales | Subjective metrics |
|---|---|
| Rathus assertiveness scale | Assertiveness |
| Connor-Davidson resilience scale; student motivation scale; resilience scale for adults | Resilience, Self-Confidence, Composure |
| Chernysenko scale | Conscientiousness |
| Big 5-factor model, Eysenck, HEXACO | Extraversion |
| Bratts Impulsiveness scale-11 | Impulsiveness |
| Situation Awareness Global Assessment Technique (SAGAT) | Situational Awareness |
| Kuder occupational interest survey | Interest |
| Motivation-perseverance-grit scale | Perseverance |
| NASA Task Load Index | Workload |

---

[2] Observer is defined as a human or equipment with methods and tools to monitor the operation, performance, and progress of an HMT and provide standard feedback to improve HMT performance.

| Scale | Pros | Cons |
|-------|------|------|
| Connor-Davidson resilience scale (CD-RISC); and student motivation scale; | The scale is well defined, and the factor analysis of this scale yielded the big five factors. The scale also demonstrates that with proper training resilience can be improved [105, 106]. | The scales focus on resilient qualities at the individual level and these scales prompt speculation. |
| Chernyshenko scale | Uses unified scores of 6 major factors, with each factor scored using analysis of 7 personality inventories, for conscientiousness computation. In-depth analysis of its effect on human performance studied in [107]. | Difficulties in assessing facets and measuring through scales due to their non-orthogonal nature. |
| Big Five-Factor model; the smaller seven; HEXACO | These models define the personality traits of a human, which have been used in designing scales for human performance as an SM. | These scales prompt the user to speculate in self-reporting. |
| Bratts Impulsiveness scale-11 | The score obtained can be used to calculate impulsiveness, which can in turn help in assessing the human performance [112]. | Self-reporting limitations that leaves room for speculations. |
| Situation Assessment Global Assessment Technique (SAGAT) | SAGAT is a well-documented tool to measure an SA, possesses a high degree of content validity based on the SA requirements analyses and is used to create the queries that were found to have predictive validity. | Limited to simulation environment most of the times. |
| Motivation-perseverance-grit scale | Grit scale enables prediction of perseverance and motivation for long-term goals. It was found to be the best predictor among many other indicators of which cadets will drop out after first difficult summer training [114]. | Self-reporting limitations that leaves room for speculations. |
| The NASA Task Load Index (NASA-TLX) | Initially introduced in 1984, efforts have been put in to make it more flexible and robust. Use of this metric since its inception is well-studied [62, 116]. | Self-reporting limitations that leaves room for speculations. |

*Decisiveness* is measured with subject ratings on the need of information, confidence in decision-making, and self-appraisal. It is also notable that peers can rate subjects' decisiveness as well [108].

*Extraversion* is measured using various rating scales such as the Big Five-Factor Model, Eysenck, and HEXACO [109, 110]. The *emotional state* of a person is calculated by ratings on behavior, facial expressions, and startle response [111]. *Impulsiveness* is measured using Bratts Impulsiveness scale-11 (BIS-11), which is the 11th version of the original 30-question inventory proposed by Bratt in 1985 [112]. *Situational awareness* is measured using the simulation technique called Situation Awareness Global Assessment Technique, which includes subjective inputs as well as objective measures [113]. *Perseverance* is measured by the scores obtained from the motivation-perseverance-grit scale that requires self-ratings [114]. *Human awareness* can be measured on a scale with the help of self or expert ratings [62, 115]. The workload is calculated using a multidimensional self-rating scale, for example, the NASA-TLX [62, 116]. Among machine metrics, self-awareness and adequacy are SM, as they require human expert ratings on deviations [61].

Table VI illustrates the pros and cons of a few popular self-reporting scales. One of the biggest drawbacks of SM is being biased in self-reported scales. For example, individuals with high neuroticism traits are expected to report more distress, pressure, etc., than others [117]. Other biases may include different socioeconomic strata, introspective ability, and image management [118, 119].

### 3) OBJECTIVE

Objective metrics (OM) are task-specific tools, functions, and formulae to measure task performance quantitatively. OM are developed to measure an activity that can be changed, customized, or expressed by a value for comparison [120]. Most identified machine and team, as well as a few human, metrics, are OM. In human metrics, *general health* can be considered an objective measure because it is measured by recording blood pressure, temperature, and heart rate [61, 121]. Similarly, *physiological fatigue* can be measured using heart rate, blood pressure, galvanic skin response, and adrenaline level. *Visual fatigue* is calibrated using Swedish occupational fatigue inventory, which employs parameters such as cardiovascular response, energy expenditure, skin temperature, and blink rate [122]. *Stress* is measured as a function of blood pressure, vocal tone, salivary alpha-amylase levels, heart rate, and blood cortisol levels [123]. *Stamina* measurement may involve taking into account parameters such physical activity (push-ups and running-speed [61]), *shift length* (the time span in which one needs to be attentive [124, 125]), or *vigilance* (through traditional human factors or modern eye-tracking methods [126]). The *memory* of an individual is measured by the degree of *recognition, relearning,* and *reconstruction* that is determined using the formulae to measure memory [127]. *Cognitive proficiency* is measured using the cognitive proficiency index, which is defined as an auxiliary scale by Wechsler intelligence scales [128].

Various *time metrics* such as *intervention response time* (time taken by the human to intervene if a problem occurs) [20], *overhead time* (time spent by the machine in idle state or unplanned activities) [80], and *productive time* (cumulative sum of time spent by the team in scheduled manual, unscheduled manual, and autonomous operations time) are also relate to objective metrics. *Neglect impart* (NI) is calculated from the NT graph by measuring the *neglect time*, or the average time before the robot's performance falls below a threshold [68]. *Settling time* is the time taken to reach the required accuracy by the machine [100]. In contrast, *completion time* is calculated for the time taken by an HMT to complete a given task. The *critical time ratio* is the ratio of the duration of the critical mission section to the duration of interaction [65]. *Task success* is calculated as the percentage of the successful tasks [80].

TABLE VII: PARAMETERS TO METRIC MAPPING

| Metric \ Parameters | blood pressure | body temperature | heart rate | galvanic skin response | adrenaline production | cortisol levels | salivary alpha-amylase | vagal tone | EEG | ECG | eye movement and blink rate | energy expenditure | overhead time | intervention response time | autonomous operations time | manual operations time | unscheduled operations time | neglect time | Settling time | time to complete | interaction frequency | true positive | false Positive | true negative | false negative | interaction time | plans level |
|---|---|---|---|---|---|---|---|---|---|---|---|---|---|---|---|---|---|---|---|---|---|---|---|---|---|---|---|
| General health | X | X | X | | | | | | | | | | | | | | | | | | | | | | | | |
| Physiological fatigue | | X | X | X | X | | | | | | | | | | | | | | | | | | | | | | |
| Visual fatigue | | | | | | | | | X | | | | | | | | | | | | | | | | | | |
| Stress | X | | X | | | X | X | X | | | | | | | | | | | | | | | | | | | |
| Productivity time | | | | | | | | | | | | | | X | X | X | X | | | | | | | | | | |
| Neglect Impart | | | | | | | | | | | | | | | | | X | X | | | | | | | | | |
| Attention allocation | | | | | | | | | | | X | | | | | | | | | | X | | | | | | |
| Mental workload | | | | X | | | | | X | | | | | | | | | | | | | | | | | | |
| Interaction effort | | | | | | | | | | | | | | | | | | | | | | | | | | X | |
| TPIR | | | | | | | | | | | | | | | | | | | | | | X | | | | X | |
| FPIR | | | | | | | | | | | | | | | | | | | | | | | X | | | X | |
| FNIR | | | | | | | | | | | | | | | | | | | | | | | | | X | X | |
| TNIR | | | | | | | | | | | | | | | | | | | | | | | | X | | X | |
| RAD/FO | | | | | | | | | | | | | | | | | | | | | | X | X | X | X | | |
| State | | | | | | | | | | | | | | | | | | | | | | | | | | | X |

The *decision-analysis* approach follows a Bayesian view of probabilities associated with the possible events, making it an objective measure [62, 129]. *Inferred mental workload* takes into account eye movement activity, cardiac functions (ECG), brain activity (EEG), and Galvanic skin response (GSR) [61]. Previously discussed metrics such as attention allocation, situation coverage, state metrics, false-alarm metrics (TP, TN, FP, and FN), RAD, and IE are also can be objective. *Human trust and reliability* on a machine are derived (i.e., inferred) from its FO factor and RAD. As the RAD increases, the user trust and reliability on the machine decreases, for example, IE, and NT inversely affect human trust and reliability. Another OM, *total coverage*, is a measure of the area or environment used by all the sensors simultaneously at a specific time during the mission execution [94]. *Neighbor overlap* can help measure how much a machine affects the performance of other machines. *Network efficiency* can be measured using bandwidth and latency.

All identified objective metrics are presented in Table VI, mapping the metrics to their corresponding parameters. Researchers can use this table to identify redundant parameters and eliminate bias.

### 4) REAL-TIME METRICS

Real-time metrics are crucial in any time-sensitive, real-time applications such as engineering, defense, and healthcare. Purposes include, but are not limited to, improving communication, response times, information transfer accuracy, and mission success rate.

*Psychomotor processing* calibrates human psychomotor speed during the mission along with *spatial processing*. These, along with stress, fatigue, general health, and various time metrics, are also known as real-time metrics. Although situation and self-awareness are crucial for a system, they cannot yet be measured in real time. Subjective measures and human attributes or traits are tough to measure in real time because they depend on the observer scale measurement, for example, adaptability, assertiveness, composure, cognitive proficiency, conscientiousness, and decisiveness. Few other metrics such as memory, decision accuracy, autonomy discrepancies, and cognitive interaction can be evaluated only after the accomplishment of the mission. Research reviewed did not indicate any of these being measured in real time even though there is a possibility of real-time measurement through recent developments in prediction models and computing. Therefore, these can be classified as non–real-time metrics [130].

### 5) SUMMARY

In general, subjective and objective measurement techniques do not measure the same parameter. However, there are a few parameters, such as cognitive load and stress, which may employ both of these measurement techniques. Based on the accuracy of the technique, one might be preferred over the other. For example, subjective measurement techniques work better on task load despite the availability of objective measurement techniques [106, 107]. Subjective metrics are recommended in combination with objective metrics for human performance. At the same time, we avoided metrics which are either derivatives of, or involve parameters similar to other metric(s). Thus, ensuring that the selected common metric set will somewhat represent those avoided metrics during the HMT performance evaluation.

Table VIII summarizes this section as a color-coded matrix representing a taxonomy in which one can look for popular metrics, relationships among metrics, selection of metric

TABLE VIII: COMPREHENSIVE COLOR-CODED CLASSIFICATION MATRIX OF HMT METRICS

| Human | | Machine | Team | |
|---|---|---|---|---|
| Adaptability | Success oriented | Self Awareness | **Cohesion** | Situation Coverage |
| Assertiveness | Psychomotor Processing | Adequacy | Cognitive Interaction | Task Allocation |
| Composure | Situation Awareness | Usability | Coordination Demand | Task Difficulty |
| Conscientiousness | Stress | Decision Accuracy | Elongation | Task Success |
| Decisiveness | General Health | Fan Out | Human Robot Ratio | Total Coverage |
| Extraversion | Human awareness | Sensory motor Coefficient | Robot Attention Demand | Similarity |
| Emotional State | Fatigue | Collision count | Critical Hazard | Autonomy Discrepancies |
| Impulsiveness | Intervention Frequency | **Plan State** | Robot Awareness | Monotonicity |
| Moral Interest | Productive Time | Network efficiency | Human Risk | Effort |
| Occupational Interest | Overhead Time | False Alarms | Critical Time Ratio | False Alarms |
| Perseverance | Reliability | Robot Adaptability | **Interventions** | FPIR |
| Reasoning | Trust | Resource Depletion | Interventions Time | FNIR |
| Resilience | Total Coverage | Interaction effort | Neglect Tolerance | Flexibility/Robustness |
| Self-Confidence | **Attention Allocation** | Mutual Delay Time | Unscheduled time | Interaction Efficiency |
| Self-Certainty | Decision Accuracy | Neglect Tolerance | **Productive time** | Network Efficiency |
| Cognitive Proficiency | **Mode Error** | Settling Time | Time Autonomous Operations | Recognition Accuracy |
| Cohesiveness | Span of Control | Time Autonomous Operations | Time in Manual Operations | Requests by Operator |
| **Judgment** | Mental workload | Time in Manual Operations | Time to complete | Requests by Robot |
| Memory | **Mental Computation** | Unscheduled Operations Time | MTBI | Team Productivity |
| Spatial Processing | Workload | **RAD** | MTCI | TNIR |
| Stamina | Guesses to Success | **Errors** | Neighbor Overlap | TPIR |
| Team oriented | Mental Models | Navigation | Plan State | Cryptic Coefficient |

Legend: ■ Performance  ■ Safety  ■ Efficiency  ■ Subjective  ■ Objective  ■ Mission  ■ Time  ■ Real Time  ■ Contact  ■ Invasive

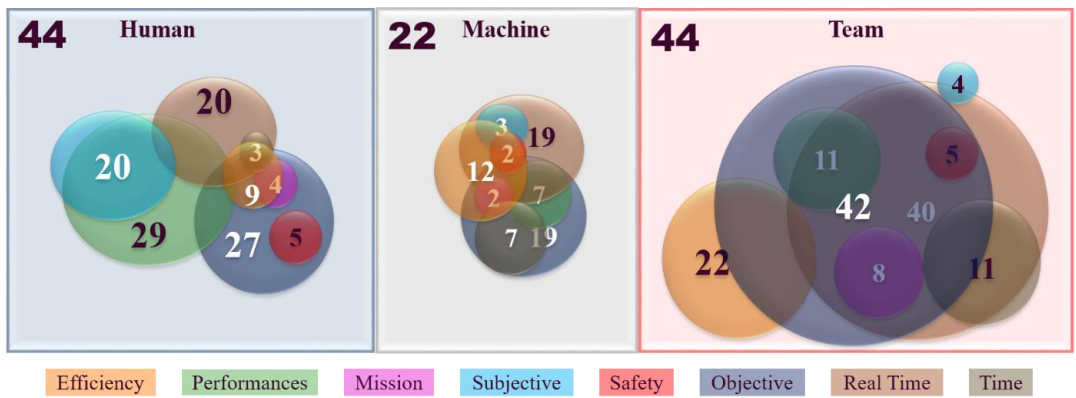

**44** Human  **22** Machine  **44** Team

Efficiency | Performances | Mission | Subjective | Safety | Objective | Real Time | Time

**Figure 4** Quantitative graphical representation of overall metrics classification (summary of Table VIII)

relevant to HMT agent, measurement techniques, and measurement aspects. Such a taxonomy is expected to allow the research community to study HMT metrics and develop a better set of common metrics. The metrics in bold are the common metrics we identify and discuss in detail in the next section. Figure 4 summarizes the color-coded table quantitatively and shows the total number of metrics represented in the table, the different aspects they measure, and measurement methods.

## IV. COMMON METRICS FOR HMT BENCHMARKING

It is understood that establishing a set of common metrics for all possible types of HMT is difficult and may not enable benchmarking for every application. Keeping that in mind, we define a set of metrics that are common to selected application areas. Nonetheless, this set may apply to a wider range of tasks or areas. Although several works attempted to identify HMT applications, our survey found only a few to either establish common metrics or at least provide guidelines for such an identification [62, 131]. Customary practices include identifying common metrics from experience, using metrics that researchers are familiar with, or attempting to measure all available aspects of a system. These approaches may lead to inefficiency due to the possible use of inappropriate measurement methods, cost of implementation, or lack of strong face validation of a measure.

In [132], researchers proposed a set of common metrics to measure the performance of interaction with the limitation of targeting only the robot or the human. Researchers identified common metrics for three agents using subjective rating scales for HRI [62, 133], which come with their share of limitations such as less performance estimation accuracy, poor reliability, spillover effects, and perspective measures (that vary based on perspective) [134-136], and which can sabotage the entire benchmarking. Another earlier work detailed a supervisory control system and proposed generalized metrics for specific examples, such as single human and HRI for multi-robot teams [137]. Later, a set of metrics for measuring supervisory control performance was selected [96]. Selection criteria of proposed common metrics are listed in Table IX. To summarize, each metric is selected based on five major aspects: total attributes a metric represents, measurement method, strong face validation of the metric, well documented in literature and practice, and supports the selected applications. Moreover, metrics must represent a team dynamics rather than an individual agent.

TABLE IX: COMMON METRIC SELECTION CRITERIA

| Items analyzed | Attributes | Selection criteria |
|---|---|---|
| Category | Safety, efficiency, time, mission and performance | Performance and efficiency (preferred if it includes others) |
| Measurement method | Subjective and/or objective, invasive and non-invasive | At least one objective method and must be non-invasive |
| Research performed | Publish research report, reviews, project data | 15+ peer reviewed publications considered |
| V&V of results | Sample size in user evaluation, accuracy of measurement, agreement in results | High sample size is preferred but multiple evaluation is mandatory |
| Application scenarios | Search, navigation, target identification, ordinance disposal, geology, surveillance, healthcare training, and tour guiding | Must be applicable in all application scenarios, measuring method can vary |

### A. COMMON HUMAN METRICS
The common metrics for human performance should give an analytical representation of human performance in an HMT. For common human metrics, we eliminated all the metrics that are invasive and only subjective to make the measurement practical. In additions to the selection criteria defined above, we focused on measurement methods for human metrics because relating activity measures to human performance is difficult [75-78]. Our research also agrees with that of several researchers in presenting trust, cognitive load, and human fatigue as important HMT metrics. However, due to lack of concrete objective measurement methods, and a strong correlation between resulting measurements and HMT performance, we excluded those from our selection. We identified four potential common human metrics: judgment, attention allocation, mental computation, and mode error.

#### 1) JUDGMENT
Judgment, or decision-making, is the process of observing and assessing situations, drawing conclusions, and predicting action consequences. It can be measured subjectively, objectively, or via mixed measurement methods. In HMT, judgment can be classified as situational or practical and may require measurement while selecting a human teammate, and in team-building and task execution, respectively. Using a combination of measurement techniques can yield a better result, including up to 90% accuracy in measuring judgment [138]. Compared to practical judgment, multiple studies have been carried in fields such as healthcare and defense to measure situational judgment. With test samples ranging from 1200 to 10600, most tests yielded accurate results [138-144].

Limitation of the method includes simulations not being representative of a practical scenario. In addition, it is known that an individual may compromise judgment for an experimental scenario and judge differently in the real world [138, 139]. Further, the *Test of Practical Judgment* [145] was found to be a prominent test for safety, social and ethical issues, and financial issues, with 134 samples showing promising results. However, the existence of only a few studies that used this method indicates a lack of widespread usage. Judgment is a mission metric that directly correlates to the human performance and efficiency in an HMT. As described above, judgment is well researched and has various studies proving the correlation with reliable results. Moreover, judgment as a mission metric represents human action in an HMT and should be able to provide the human factor analysis needed in HMT benchmarking.

#### 2) ATTENTION ALLOCATION
In stressful situations with complex systems, it is possible that focus is shifted from an important task to a minor or an unimportant task [69]. Therefore, it is expected that tracking real-time attention allocation will improve an HMT. A 2008 review discussed a few attention metrics including eye tracking, verbal protocols, and tracking resource allocation cognitive strategies (TRACS) [61]. Several studies were performed in TRACS, with a maximum participant size of up to 45, showing a correlation with attention allocation. TRACS is achieved by measuring HCI with a 2D representation of a human [146, 147], and a common limitation involves customization for each interface and task [61]. Various researchers have studied and correlated eye tracking, attention allocation, and human performance using fixations, saccades, pupillometry, and blinks for application areas such as UAVs, supervisory control, and healthcare [61, 148-151], while encountering limitations such as limited correlation between gaze and thinking, intensive data analysis, and noise in the measured data [152]. In conclusion, an effective measure of attention can be achieved through combining eye tracking and TRACS. As described above, attention allocation is a well-studied metric that has different measurement methods and satisfied the criteria to be selected as a common metric. In addition, it is noteworthy that attention allocation deals with human parameters that directly affect HMT performance.

#### 3) MENTAL COMPUTATION
Mental computation, mental workload, and cognitive load are well-studied theories and recent studies establish their correlation with human performance [153], satisfying our criteria of common metric selection. However, since mental computation is a non–real-time metric, HMT developers need to perform mental computational studies and adjust their design for peak performance. Primary measurement methods use subjective scaling and physiological performance parameters. Performance measures can be used to measure relative speed, accuracy, and elapsed time [153]. Physiology studies involving mental computation and human performance are overwhelming, as studies include EEG, ECG, GSR, eye tracking, etc., with participants ranging from 28 to 300 and tasks ranging from defense and medicine to controls [154-159]. Studies successfully differentiated between multiple and increased mental workload based on task demand but failed to show a consistent correlation between efficiency and identified cognitive load patterns. Therefore, these measurement methods should be used only for minimizing mental workload at the HMT design stage, which may result in better performance during operation.

### 4) HUMAN ERROR
An error has a direct correlation with performance and efficiency. Every system needs to rectify mistakes if any in real-time and post-mission completion. Along with the previously mentioned selection criteria, the above two factors led to the selection of human error as a common metric. This category of error is one of the most prominent and important metrics. There are several types of errors presented in the literature. However, *mode error* is one of the most studied metrics. Mode error represents the human error that affects HMT operation. Mode error is defined as the difference in actual and intended operation mode as a result of either a human-machine miscommunication or a human selecting an incorrect mode of operation [99]. Mode error is a widely studied and prominent human error that can affect the human-machine relationship and depends on the application scenario. Mode error can adversely affect performance based on the severity of the error. If unchecked, a mode error may result in total system failure. Researchers have measured mode error during system operation in various ways [99, 160, 161]; for example, mode of operation must change when flying conditions change while flying a single-engine airplane with focus on airspeed, altitude, and routing by controlling thrust, ailerons, elevators, and rudder. Otherwise, a mode error occurs and might leads to catastrophic system failure. Mode error can be converted to an empirical value for some applications. However, it is noteworthy that all scenarios cannot be easily generalized.

### B. COMMON MACHINE METRICS
To select a common metric from the identified metrics, an application-specific primary analysis was conducted. A machine parameter can be easily measured; however, identifying a metric that may apply in broad application space is quite challenging. In this section, we identified metrics that have a maximum number of mutually exclusive parameters in addition to the selection criteria defined in Table IX. The goal is to provide metrics that measure performance, efficiency, and accuracy of task operations while minimizing parameter redundancy. We have identified three potential common machine metrics described below.

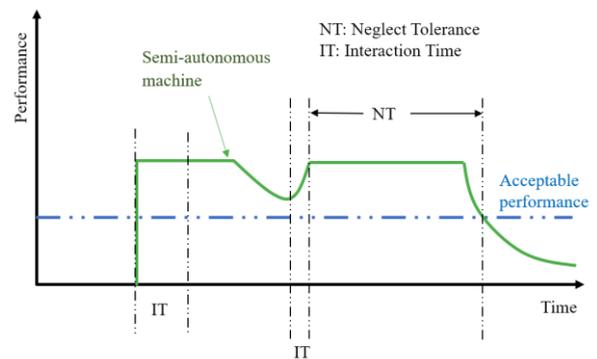

**Figure 5 Neglect Tolerance (NT) Model**

### 1) ROBOT ATTENTION DEMAND (RAD)
RAD represents the relationship of the machine with human teammate and is measured using NT and IE, as shown in equation 1 [132, 162, 163]. We further discuss these in detail.

$$RAD = \frac{IE}{IE+NT} \qquad (1)$$

- **Neglect tolerance (NT)** is a unique characteristic graph of machine performance that is measured for each autonomous system individually, as shown in figure 5 [132, 164]. NT usually follows a decreasing trend with time, while the rate of change varies from machine to machine [132, 162]. Although no standards have been established or adopted for NT measurement, several researchers have adopted NT in their studies [122,153,155].
- **Interaction effort (IE)** is the capability of the machine to understand the higher human communication level. It is not just a physical input or account for stages of understanding information and decision-making; it can be inferred from secondary parameters. For example, eye tracking can be used to determine whether a human was looking at the display before an input. Therefore, including time for these tasks would be more accurate. A hypothetical interaction effort characteristic graph is one of the prominent models many researchers have adapted, where IE is estimated using RAD and interaction time [162]. However, in practice, researchers have measured the interaction time as the IE [163].

In relating RAD to autonomous systems' performance, studies state [132] and experimentally found that a lower RAD value results in a better performance [162, 163]. Most of the experiments conducted in this area are related to robots and software agents. The metric has not been evaluated with humans or an HMT. RAD is another well-documented metric in HMT literature that can measure machine performance and real-time efficiency while the machine is operating in an HMT setting. RAD is a unique metric developed solely for machine teammates in an HMT. Even though primary results satisfy the selection criteria for RAD, several mathematical models suggest further possible development.

### 2) MACHINE STATE METRIC

Machine state metric was coined and first used for an airplane in 2010 [80]. However, measuring or identifying a machine state and its changes is a widespread practice. Possible states for a given autonomous system are represented as a state chart, and popular types include the rendezvous manager state chart (four states), data flow manager state chart (five states), and unified modeling language statechart (varying number of states) [165-167]. State measurement is helpful in real-time system observation and correlating machine clock time and machine performance. Machine state can also provide a sense of a machine's operation level and facilitate monitoring. Following is an example of how the state of a machine could be measured. For example, if a machine has four states—assigned, executed, idle, and out of plan—only one state can be true at a single point in time, and enumerated task status values are indicated as the following: failure = >0, successful = 0, executing = −1, paused = −2, and pending = −3 [80], then:

$$State\ at\ time\ t = \begin{cases} plan\ assign\ status = -3 \\ plan\ execution\ status = -1 \\ plan\ idle\ state = -2 \\ out\ of\ plan\ status \geq 0 \end{cases}$$

### 3) ERRORS

The error is one of the prominent metrics that can be represented in several ways based on the machine and application type. An error has a direct correlation with machine performance, efficiency, and task success. Machine errors also affect the team performance in several ways, ranging from affecting user trust to increasing workload, stress, and fatigue. If the errors are high and frequent, creating an HMT would be counterproductive to the mission or task. Machine errors include various types of faults and defects and vary based on the application [168]. For example, hardware fault in a vehicle may lead to a hardware error, while *interpretation error* appears due to the environmental conditions, which are difficult to model [169]. Researchers have also described a few other errors related to software intelligent assistants such as interaction errors, data entry errors, cumulative calculation errors, cognitive overload, misrepresentation of information, and security errors [170, 171]. Error correction methods rely on error types, and the popular methods include simulation, modeling, testing, verification, and validation; for example, modeling errors can be avoided using simulation while hardware faults are identified through verification during design and development [169, 172, 173]. There is no empirical formulation for errors, in general; however, based on application and error type, a unique value can be awarded to an error to represent the effect of the error on performance.

### C. COMMON Team METRICS

After careful analysis, we identified three potential common team metrics that can be used in measuring the performance of a team or system that satisfies the selection criteria established in Table IX. The combination of these three metrics, along with human and machine metrics, will provide an overall HMT performance score level and are discussed below:

### 1) PRODUCTIVE TIME

Measuring time is a relatively simpler and more reliable way to achieve higher accuracy when compared with other metric measurement techniques. Productive time is a metric that is used widely to measure team productivity by measuring the time spent by the machine and the human on a mission, and it is represented by the following equation:

$$\begin{aligned} Productive\ &time \\ &= \textstyle\sum Autonomous\ operation\ time \\ &+\ Manual\ operation\ time \\ &+\ Unscheduled\ manual\ operation\ time \end{aligned}$$

Many researchers have used productive time and total task time to measure their robot and teams' performance [8, 80, 174, 175]. For example, in a task of transferring an object from location A to location B, productive time involves object retrieval time, travel time, and replacement time. Other times such as planning, rerouting, and delays in communication will be added to task completion time but not productive time. Productivity is calculated as the ratio of productive time and total task time. Productive time evaluates team productivity and efficiency, which is a key parameter defining the team success. It is also a well-studied metric, measurable in real time, and can be objective with stronger face value. All of these contribute to its selection as a common metric.

### 2) COHESION

Cohesion is defined as a dynamic process that is reflected in the tendency of a group to remain united in the pursuit of its instrumental objectives and for the satisfaction of member affective needs [176, 177]. Our survey identified hundreds of published research studies and books on cohesion in human teams, indicating its importance in team performance evaluation. We have selected it as a common metric based on our review of both human team studies and HMT studies because of cohesion's strong and direct relation to team performance. It also satisfies all the requirements established in Table IX and represents the effect of a team on each HMT agent. Cohesion has been studied in human teams to improve team performance since 1978. The group environment questionnaire is a widely used self-reporting subjective method for measuring cohesion [177]. According to a review, cohesion demonstrated a significant effect on team performance [178] and can be measured unidimensionally or multidimensionally, with the latter being better. As an SM, it is difficult to incorporate it in real time. However, team measurements prove that cohesion is a function of time and plays a key role in measuring the extent to which a team can work together before deploying a team. A standard method of measuring cohesion in HMTs has still to be developed. Several researchers used communication patterns between teammates and connected members to measure cohesion [94, 179, 180]. In conclusion, cohesion plays a key role as a metric to measure team performance or teaming nature.

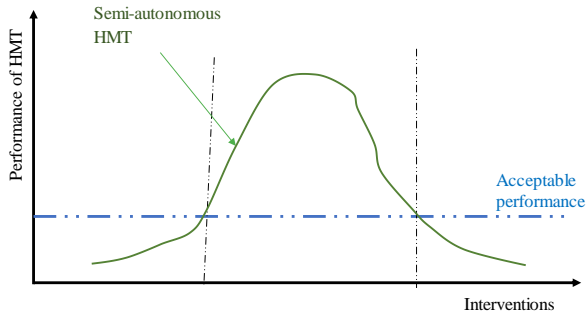

**Figure 6** Hypothetical interventions curve

### 3) INTERVENTIONS

Although human intervention may have a negative impact on the overall HMT performance, it is necessary to resolve errors made by a machine. It has been used widely to represent autonomous system performance and correlates the number of interventions to the performance of a machine [181]. Timely and optimal number of interventions by a teammate will lead to better performance in an HMT [182]. It can be measured in intervention time or intervention response time, which is measured as the total time a machine spent responding to interventions. In both methods, simple timers or counters can be used [62, 183]. However, this metric needs more in-depth studies before a standard to measure HMT performance can be developed. Based on analyzed studies, we hypothesize that intervention is a nonlinear function with an inverted U-shaped dose-effect curve drawn against a performance of an HMT as shown in figure 6. For example, too many interactions in a synthetic assistant-based learning environment may cause interruptions in learning while too few may give reasons to repeat earlier mistakes.

### D. DISCUSSION

Selected common metrics in all three agents of HMT may be helpful to measure the HMT performance and derive an empirical value to allow comparison with another HMT [184]. This measurement is multidimensional (application, scenarios, agent, etc.) and will give an in-depth analysis of difficulties in HMT applications that need to be improved to achieve better performance. We have identified 10 common metrics among more than 100. These metrics have many parameters as sub-metrics that allowed detailed HMT analysis for aspects such as safety, performance, and efficiency. In [131], researchers attempted to analyze the possibility of a machine as a teammate. Their concluding remark suggests that future HMT researchers may need to identify the uniqueness of a machine and design an HMT such that members (human/robot) complement each other rather than designing a system in which a robot merely imitates a human. In this context, selected common metrics may help measure each agent individually, measure the HMT independently, and help future engineers design tailored HMTs and benchmarks. Through this study, we found that the performance of the HMT is rated based on a performance-score, which is a weighted combination of common metrics [96, 184, 185]. This score can act as an application-specific HTM benchmark and provide a relative performance score, thus providing a platform for HMT comparison [186, 187].

## V. CONCLUSION

Synthetic teammates are moving from personal voice assistants that answer questions such as "How's the weather outside?" and set meeting reminders to assistants that can be used in healthcare, large-scale industrial production systems, military surveillance, threat neutralization, and national security. These application areas typically entail a threat to the human life along with huge investments that make use of a standardized HMT essential for task execution. To conclude, we would like to point out the importance of an application-specific HMT performance evaluation that could use the identified common metrics.

TABLE X: SUMMARY OF IDENTIFIED COMMON METRICS

| Metric | Agent | Selection criteria | Description |
|---|---|---|---|
| Judgement | Human | Objective, non-real-time, user studied, reviews, and correlation with team performance | Measures human judgement skills and trust levels, can be measured at design or application stage |
| Attention allocation | Human | Objective, real-time, user studies, reviews, and correlation with human performance | Proven measurement techniques will measure human attention allocation efficiency that can be related to human performance |
| Mental computation | Human | Objective and/or subjective, user studies, reviews and correlation with HMT design | Using EEG techniques, we can create human mental models that will be useful in HMT design and development |
| Human Error | Human | Objective, real-time, published results, relation with task execution | Real-time mode error measurement will help HMT execute its tasks with precision |
| RAD | Machine | Objective, real-time, published results, characteristic graphs and relation with machines performance | RAD monitoring will provide significant results that can be used to measure machine performance |
| State | Machine | Objective, real-time, published results, characteristic graphs and relation with machines performance | Machine state can be used by human to understand the machine, and help improve machine and team performance |
| Errors | Machine | Objective, real-time, published results, characteristic graphs and relation with machines performance | Being a generalized metric that gives all machine errors as a quantitative value and can be used in performance evaluation |
| Productive time | Team | Objective, real-time, published results, characteristic graphs and relation with team performance | Being a time metric, it can be used to significantly identify team success |
| Cohesion | Team | Objective, real-time, published results, characteristic graphs and relation with team performance | An observer metric and helps in identifying HMT teaming nature quantitatively |
| Interventions | Team | Objective, real-time, published results, characteristic graphs and relation with team performance | Can be positive or negative in performance score formula and plays a crucial role in understanding team |

Through this review, we proposed a definition and identified the components and functional blocks of an HMT. At the beginning of the review, we posed three goals to achieve through this review: (i) identify available metrics in HMT, (ii) analyze and classify identified metrics, and (iii) propose common metrics. Available metrics were identified in section 3.1; analysis and classification were achieved in section 3.2; and finally, we proposed 10 common metrics to evaluate HMTs in section 4. The common metrics have also been summarized in Table X. Metric versus parameter table, and color-coded metrics table are ancillary results of the review.

Although a common metric may be used for various applications, the interpretation of the scores might be application specific; for example, a UAV HMT will have a different scoring mechanism than a healthcare assistant. In conclusion, selecting appropriate test populations when benchmarking an HMT is very important. Specifically, as robots are increasingly deployed in applications where the target user is not an expert roboticist [188], it becomes critical to recruiting subjects with a broad range of knowledge, experience, and expertise. The continuing work under this effort will expand and refine the material presented here. The eventual plan is to provide a living, comprehensive document that future research and development efforts can utilize as an HMT metric toolkit and reference source.

## ACKNOWLEDGMENT

This work was partially supported by the Electrical Engineering and Computer Science Department, Paul A. Hotmer Cybersecurity and Teaming Research (CSTAR) Lab at the University of Toledo, and Grant award "Improving Healthcare Training and Decision Making Through LVC" from the Ohio Federal Research Jobs Commission (OFMJC) through Ohio Federal Research Network (OFRN). The team would like to thank Colin Elkin, Ruthwik Junuthula and Mayukha Cheekati for their help in proofreading and editing.

**PRAVEEN DAMACHARLA** (S'11–GS'13) received the B.Tech., degree in electrical and electronics engineering from the Koneru Lakshmaiah College of Engineering affiliated to Acharya Nagarjuna University, Guntur, A.P., India, in 2012, and is currently working towards his Ph.D. degree at the University of Toledo, Toledo, OH, USA. He is currently a Research Assistant with the Department of Electrical Engineering and Computer Science, University of Toledo. His research interests include Human-Machine Teaming, Human factors, Machine learning, Autonomous synthetic assistants and applied robotics. Mr. Damacharla was the recipient of Outstanding Teaching Assistant 2015 award presented by College of Engineering, The University of Toledo. He was also the recipient of the 2016 Advanced Leadership Academy scholarship from College of Business and Innovation, University of Toledo.

**AHMAD Y. JAVAID** (GS'12, M'15) received his B.Tech. (Hons.) Degree in Computer Engineering from Aligarh Muslim University, India in 2008. He received his Ph.D. degree from The University of Toledo in 2015 along with the prestigious University Fellowship Award. Previously, he worked for two years as Scientist Fellow in Ministry of Science & Technology, Government of India. He joined the EECS Department as an Assistant Professor in Fall 2015 and is the founding director of the Paul A. Hotmer Cybersecurity and Teaming Research (CSTAR) lab. His research expertise is in the area of cybersecurity of drone networks, smartphones, wireless sensor networks, and other systems. He is also conducting extensive research on human-machine teams and applications of AI and machine learning to attack detection and mitigation. During his time at UT, he has participated in several collaborative research proposals that have been funded by agencies including the NSF (National Science Foundation), AFRL (Air Force Research Lab), and the State of Ohio. He has published more than 50 peer-reviewed journal, conference, and poster papers along with several book chapters. He has also served as a reviewer for several high impact IEEE journals and as a member of the technical program committee for several conferences.

**JENNIE J. GALLIMORE** is a Professor of Industrial and Human Factors Engineering in the Department of Biomedical, Industrial and Human Factors Engineering at Wright State University. She also holds a Joint Appointment as a Professor in the Department of Surgery in the Boonshoft School of Medicine. She is the Associate Dean of the College of Engineering and Computer Science. She received her Ph.D. in Industrial Engineering and Operations Research, from Virginia Polytechnic Institute and State University (1989) and holds a Master's degree in Psychology. Dr. Gallimore applies human factors and cognitive engineering principles to the design of complex systems. She has worked in the research domains of visualization of information, aviation, unmanned aerial systems, health care systems, petrochemical, and virtual environments to name a few. Dr. Gallimore has published over 60 technical articles and has taught 13 different courses related to human factors engineering. She is a recipient of the 2001 AIAA Simulation and Modeling Best Paper Award, and the WSU College of Engineering and Computer Science Excellence in Research Award 2001/2002. Three of her Ph.D. students received the Stanley N. Roscoe award for best dissertation in Aviation Human Factors (2000, 2003, 2007).

**VIJAY K. DEVABHAKTUNI.** (S'97–M'03–SM'09) received the B.Eng. degree in electronics and electrical engineering and the M.Sc. degree in physics from the Birla Institute of Technology Science, Pilani, India, in 1996, and the Ph.D. degree in electronics from Carleton University, Ottawa, ON, Canada, in 2003. He held the Natural Sciences and Engineering Research Council of Canada Post-Doctoral Fellowship and spent the tenure researching with the University of Calgary, Calgary, AB, Canada, from 2003 to 2004. From 2005 to 2008, he held the internationally prestigious Canada Research Chair in computer-aided high-frequency modeling and design with Concordia University, Montreal, QC, Canada. In 2008, he joined the Department of Electrical Engineering and Computer Sciences, The University of Toledo, Toledo, OH, USA, as an Associate Professor. Since 2012, he has been the Director of the College of Engineering for Interdisciplinary Research Initiatives, Toledo, OH, USA, where he has been recently promoted to a Full Professor. He has co-authored around 190 peer-reviewed papers. His current research interests include applied electromagnetics, biomedical applications of wireless networks, computer-aided design, device modeling, image processing, infrastructure monitoring, neural networks, and RF/microwave optimization. Dr. Devabhaktuni serves as an Associate Editor of the International Journal of RF and Microwave Computer-Aided Engineering